\documentclass[11pt,a4paper,oneside]{article}
\usepackage[top=1in,bottom=1in,left=.5in,right=.5in]{geometry}
\usepackage[margin=.5in,small]{caption}

\usepackage{authblk}
\usepackage{cite}
\usepackage{amsmath,gensymb}
\usepackage{epsfig}
\usepackage{graphics}
\usepackage{amsfonts, amssymb, upgreek}
\usepackage{textcomp}
\usepackage[english]{babel}
\usepackage{blindtext}
\usepackage{color}

\providecommand{\abs}[1]{\vert #1\vert}

\providecommand{\srho}[1]{\hspace{.1ex}\varrho_{_{#1}}}

\newcommand{\bsigma}{\boldsymbol{\sigma}}
\newcommand{\bvarrho}{\boldsymbol{\varrho}}
\newcommand{\balpha}{\boldsymbol{\alpha}}
\newcommand{\bx}{\boldsymbol{x}}
\newcommand{\by}{\boldsymbol{y}}
\newcommand{\bu}{\boldsymbol{u}}
\newcommand{\bH}{\boldsymbol{H}}

\newcommand{\cZ}{\mathcal{Z}}
\newcommand{\cZa}{\mathcal{Z}_\mathrm{a}}
\newcommand{\cZs}{\mathcal{Z}_\mathrm{s}}

\newcommand{\vp}{\varrho_{_{+}}}

\newcommand{\vm}{\varrho_{_{-}}}

\newcommand{\vpm}{\varrho_{_{\pm}}}

\newcommand{\Pst}{P\!_\mathrm{st}}

\newcommand{\upd}{\mathop{}\!\mathrm{d}}
\newcommand{\erf}{\mathrm{erf}}
\newcommand{\Or}{\mathrm{O}}
\begin{document}

\title{Revisiting random deposition with surface relaxation: approaches from growth rules to Edwards-Wilkinson equation}

\author[1,2]{R. C. Buceta\thanks{rbuceta@mdp.edu.ar}}
\author[1]{D. Hansmann\thanks{David.Hansmann@conicet.gov.ar}}
\author[1]{B. von Haeften}

\affil[1]{Departamento de F\'{\i}sica, FCEyN, Universidad Nacional de Mar del Plata}
\affil[2]{Instituto de Investigaciones F\'{\i}sicas de Mar del Plata, UNMdP and CONICET}
\affil[{ }]{Funes 3350, B7602AYL Mar del Plata, Argentina}
%\date{Received: date / Revised version: date}

\maketitle

\abstract{
We present several approaches for deriving the coarse-grained continuous Langevin equation (or Edwards-Wilkinson equation) from a random deposition with surface relaxation (RDSR) model. First we introduce a novel procedure to divide the first transition moment into the three fundamental processes involved: deposition, diffusion and volume conservation. We show how the diffusion process is related to antisymmetric contribution and the  volume conservation process is related to symmetric contribution, which renormalizes to zero in the coarse-grained limit. In another approach, we find the coefficients of the continuous Langevin equation, by regularizing the discrete Langevin equation. Finally, in a third approach, we derive these coefficients from the set of test functions supported by the stationary probability density function (SPDF) of the discrete model. The applicability of the used approaches to other discrete random deposition models with instantaneous relaxation to a neighboring site is discussed.      
}

%\begin{keyword}
%Surfaces, interfaces, growth processes
%\end{keyword}

%\pacs{02.50.-r, 05.10.Gg, 68.35.Ct, 89.75.Da}

%\begin{keyword} random deposition with surface relaxation model \sep lattice model \sep growth processes \sep coarse-grained approach \sep continuous Langevin equation \sep Edwards-Wilkinson equation \sep generalized functions \sep distributions \sep traslations \sep test space \end{keyword}

\section{Introduction}
Discrete models and continuous Langevin equations for surface growth are usually grouped into universality classes. Each universality class is distinguish by  a set of power-law exponents of statistical observables related to each other by scaling relations \cite{Barabasi-95}. These exponents are frequently determined by Monte Carlo simulations \cite{Family-88}. Models and equations within the same class share not only the same scaling exponents, but also the same linearities and (or) nonlinearities \cite{Lai-91}. There are two theoretical methods to connect discrete models to the corresponding continuous equations within a universality class. Both formalisms are based on the discrete Langevin equation obtained by the Kramers-Moyal expansion of the master equation.  In both approaches, the first and second transition moment of the discrete Langevin equation are interpreted as generalized functions (or distributions) evaluated at the lattice points. The transition moments, which determine the behaviour of system, are normally expressed in term of if-them-else structures. Consequently, the transition moments could be a proper combination a Heaviside (or unit-step) generalized functions. The first of the two approaches was introduced by Vvedensky {\sl et al.} \cite{Vvedensky-93} and is based in the regularization of the transition moments by means of the individual regularization of each Heaviside included within the moments. The regularized Heaviside functions are sigmoid curves ({\sl e.g.} trigonometric, hyperbolic, tangent, or error functions) that allow Taylor expansions. The standard procedure is to replace each Heaviside function $\Theta$ by a regularizing function $\theta_\varepsilon$ which admits a Taylor expansion around zero \cite{Park-95,Costanza-97,Buceta-05,Muraca-04}. Here, $\theta_\varepsilon$ depends on a continuous regularization parameter $\varepsilon$, in a way that $\theta_\varepsilon\to \Theta$ when $\varepsilon\to 0$. The Taylor coefficients depend on both the regularization function and the regularization parameter \cite{Hansmann-13}. It is possible to obtain the continuous Langevin equation of the corresponding universality class by first expanding the regularized first transition moment around zero and then coarse-graining the expression. The second approach was introduced by Buceta and Hansmann \cite{Buceta-12}. In contrast to Vvedensky's approach, it employs the transition moments as generalized functions, without further regularization. Buceta and Hansmann applied these generalized functions to a set of test functions supported by the SPDF of the discrete process. This theoretical approach is inspired by the surface tilting method which is used to determine  {\sl via} Monte Carlo simulations the nonlinearities of the discrete models ({\sl e.g.} models belonging the Kardar-Parisi-Zhang (KPZ) universality class). Using this approach we can derive all coefficients, the linear and  nonlinear coefficients as well as the noise intensity of the continuous Langevin equation, which characterizes the universality class of the underlying discrete model. To that end, the method uses small transformations in the test space to find the coarse-grained coefficients of restricted and unrestricted discrete processes, {\sl e.g.} the restricted solid-on-solid model and the ballistic model, respectively. 

The lattice growth models with random deposition followed by instantaneous relaxation to a neighboring site show transition rules which contain three basic microscopic processes: deposition, diffusion, and volume conservation. The vertical incoming flux of particles on the substratum is the source of non-conserved noise in the system. In contrast, systems without particle inflow show conserved noise \cite{Hansmann-13}. The surface relaxation rules lead to microscopic processes of diffusion and volume conservation. However, in the coarse-grained limit these models belong to one of three groups (universality classes); models which show only diffusion, only volume conservation, or both diffusion and volume conservation. The RDSR model (or Family model) belongs to the first group, the Edward-Wilkinson universality class, and shows only deposition with diffusion process, while volume conservation processes renormalizes to zero. In the second group one finds the Das Sarma-Tamborenea (DST) molecular-beam-epitaxy (MBE) growth model which belongs to the Villain-Lai-Das Sarma (VLDS) universality class. This class is define by the VLDS equation, which includes deposition with volume conservation processes. In the third group one finds an intermediate model, the Wolf-Villain MBE growth model, which shows a crossover from the VLDS universality class to the EW class. For the three groups the generic continuous Langevin equation with non-conserved noise, for the height $h=h(x,t)$, is
\begin{equation}
\frac{\partial h}{\partial t}=F_0+\nabla^2 Z_\mathrm{s}+\nabla\!\cdot\! Z_\mathrm{a} +\eta(x,t)\;,\label{cLE}
\end{equation}
where $F_0$ is the incoming flux, $Z_\mathrm{s}$ is known as symmetric (or conserved KPZ) kernel, and $Z_\mathrm{a}$ is known as antisymmetric kernel. Both kernels are functions of $\nabla h$ and/or $\nabla^2 h$. The term $\nabla^2 Z_\mathrm{s}$ describes the volume conservation process and the term $\nabla\!\cdot\! Z_\mathrm{a}$ describes the diffusion. The mean value of the non-conservative noise $\eta$ is zero, {\sl i.e.} $\langle\eta(x,t)\rangle=0$, and correlation is
\begin{equation}
\langle\eta(x,t)\,\eta(x',t')\rangle=2\,D\,\delta(x-x')\,\delta(t-t')\;,\label{corr-noise}
\end{equation} 
where $D$ is the noise intensity. Before going to the coarse-grained limit, the symmetric kernel of the RDSR model renormalizes to zero ({\sl i.e.} $Z_\mathrm{s}= 0$) and the antisymmetric kernel is
\begin{equation}
Z_\mathrm{a}=\left[\nu_0+\frac{\nu_2}{3}\,\abs{\nabla h}^2\right]\,\nabla h\;,\label{kernel-Za}\end{equation}
up to the second-order of $\abs{\nabla h}$, that leads to the inhomogeneous diffusion equation 
\begin{equation}
\frac{\partial h}{\partial t}=F_0+\nu\,\nabla^2 h+\eta\;,\label{IDe}
\end{equation}
with the diffusion term $\nu=\nu_0+\nu_2\,\abs{\nabla h}^2+\mathsf{O}(4)$. The antisymmetric kernel of the DST MBE growth model renormalizes to zero ({\sl i.e.} $Z_\mathrm{a}= 0$) and the symmetric kernel is 
\begin{equation}
Z_\mathrm{s}=\mu\,\nabla^2 h+\frac{\lambda}{2}\,\abs{\nabla h}^2\;,\label{kernel-Zs}
\end{equation}
in the lowest order in Laplacians and gradients. The equation with $\mu<0$ and $\lambda<0$  leads to the VLDS equation \cite{Villain-91,Lai-91} and with $\mu<0$ and $\lambda=0$ it leads to the stochastic Mullins-Herring equation \cite{Herring-51,Mullins-57}. 

In order to obtain the EW equation \cite{Edwards-82} from eq.~(\ref{IDe}) it is common to introduce the coarse-grained space and time variables through the replacements $x\to b^{-1}x\,$, $t\to b^{-z}t\,$, where $z$ is a proper scaling parameter and $b$ parametrizes the extent of coarse graining in such way that the continuum limit is recovered with $b\to 0$. The corresponding coarse-grained height function is $u(x,t)=b^{\alpha}(h-F_0\,t)\,$, where $\alpha$ is another proper scaling parameter. The coarse-grained noise is given by $\zeta=b^{-(1+z)/2}\,\eta$. Defined in this way, it leave the noise correlation intensity of eq.~(\ref{corr-noise}) invariant under the coarse-grained transformation. Using these replacements in eq.~(\ref{IDe}) one obtains
\begin{equation}
\frac{\partial u}{\partial t}=b^{2-z}\nu\,\nabla^2 u+ b^{\alpha-(z-1)/2}\,\zeta\;,\nonumber
\end{equation}
with $\nu=\nu_0+b^{2(1-\alpha)}\nu_2\,\abs{\nabla u}^2+\mathsf{O}(b^{4(1-\alpha)})$. Setting $z=2\;,\alpha=1/2$ and taking the limit $b\to 0$, the diffusion coefficient $\nu\to\nu_0$ and one obtains the EW equation
\begin{equation}
\frac{\partial u}{\partial t}=\nu_0\,\nabla^2 u+ \zeta\;.\label{EWe}
\end{equation} 
In this paper we review the RDSR model introduced by Family \cite{Family-86}. We start linking their growth rules to the Edwards-Wilkinson equation. In Section~\ref{sec:RDSR} we show that its first and second transition moment are separable into three different processes: deposition, diffusion, and volume conservation. Doing so we show that the diffusion (volume conservation) process is related to antisymmetric (symmetric) kernel, which are the base to simplify the derivation of the continuous equation or its coefficients in the following sections. In Section~\ref{sec:regulariz} we derive the continuous Langevin equation {\sl via} regularization of both kernels using the known method of $\Theta$-function regularization \cite{Vvedensky-93}. We show that the symmetric kernel renormalizes to zero in the coarse-grained limit and that the remaining antisymmetric kernel can be identified as the diffusion term of the EW equation.  In Section~\ref{sec:framework} we include an overview on generalized function theory applied to the study of discrete surface models. We show how to derive the coefficients of the continuous Langevin equation in terms of the set of test functions supported by the SPDF of the discrete model. In Section~\ref{sec:coef-test} we determine the coefficients {\sl via} small translations in the test space. At the end of this section, using the separability of the test functions in polar coordinates, we find the coarse-grained coefficients in terms of radial functions. Finally, a summary is given in Section~\ref{sec:concl}.

\section{RDSR model: from discrete to continuous\label{sec:RDSR}}
%\subsection{Family Model}

All discrete models with random deposition followed by instantaneous relaxation to a neighbouring site can be described similarly. The incoming flux of particles falls sequentially toward the substratum at randomly chosen columns. According to the rules, the incoming particles try to stick to the columns or, otherwise, relax to its lowest neighbouring column. The time necessary to deposit a particle layer of the average size $a$ is $\tau$. We assume a discrete process that takes place on a square lattice of length $L$ with the unit cell size $a$. Furthermore, it is assumed that a randomly chosen column or some next neighbouring (NN) column can grow only $a$. The surface configuration $\bH$ of the system is determined by the set of heights $\{h_j\}$ which corresponds to the columns $j=1,\dots ,N$ (with $L=N a$). The transition rate $W(\bH,\bH')$ of a RDSR model, which describes the change between two consecutive surface configurations $\bH$ and $\bH'$ in the lapse $\tau$ is 
\begin{eqnarray} 
&&W(\bH,\bH')=\frac{1}{\tau}\,\sum_{k=1}^N\Bigl[w_k^{(0)}\,\Delta(h'_{k}-h_{k}-a)\,\prod_{j\neq k}\Delta(h'_j-h_j)+w_k^{(-)}\,\Delta(h'_{k-1}-h_{k-1}-a)\!\prod_{j\neq k-1}\Delta(h'_j-h_j)\nonumber\\&&\hspace{21ex}+\,w_k^{(+)}\,\Delta(h'_{k+1}-h_{k+1}-a)\!\prod_{j\neq k+1}\Delta(h'_j-h_j)\Bigr]\;,\label{transition}
\end{eqnarray}
where $\omega_k^{(\cdot)}$ is the probability at which the deposition process occurs when the chosen column is $k$. The superscript labels {\scriptsize$(\pm)$} denote deposition by relaxation to one of the neighbouring columns of the column $k$. The superscript label {\scriptsize$(0)$} denote deposition at the column $k$. Here $\Delta(z)$ is equal to 1 if $z=0$ and equal to $0$ otherwise. The first transition moment is
\begin{equation} K_j^{(1)} =\sum_{\bH'}(h'_j-h_j)\,W(\bH,\bH')=\frac{a}{\tau}\left[w_{j-1}^{(+)}+w_j^{(0)}+w_{j+1}^{(-)}\right]\;,\label{1st-transition}
\end{equation}
and the second transition moment is
\begin{equation}
K_{ij}^{(2)} = \sum_{\bH'}(h'_i-h_i)(h'_j-h_j)\,W(\bH,\bH')= a K_j^{(1)}\, \delta_{ij}\,.\label{2nd-transition}
\end{equation}
The growth rules of the RDSR (or Family) model are the following. A particle is deposited on the substrate at the chosen column if the neighbouring columns are not lower in height. Otherwise, the particle is deposited on the substrate at the neighbouring column with lower height or in case of equal heights at one randomly chosen neighbouring column. The probabilities of the RDSR model are
\begin{eqnarray}
w_j^{(0)}&=&\Theta(h_{j+1}-h_j)\,\Theta(h_{j-1}-h_j)\;,\label{dd}\\
w_{j\pm 1}^{(\mp)}&=&\tfrac{1}{2}\left[1+\Theta(h_{j\pm 2}-h_{j\pm 1})\right]\left[1-\Theta(h_j-h_{j\pm 1})\right]\;,\label{rd}
\end{eqnarray}
where $\Theta$ is the Heaviside (unit step) function which is defined as $\Theta(x)=1$ if $x\ge 0$ and $\Theta(x)=0$ if $x<0$. As mention above, the probability $w_j^{(0)}$ refers to the deposition at the chosen column $j$. In contrast, the probabilities $w_{j\pm 1}^{(\mp)}$ refers to the relaxation to column $j$ when the chosen column is $j\pm 1$. From Eqs. (\ref{dd}) and (\ref{rd}) we obtain the identity $w_j^{(0)}+w_j^{(-)}+w_j^{(+)}=1$ which ensures that the deposition average rate is $1/\tau$. Following the master equation approach \cite{Kramers-40} the discrete Langevin equation, for each height $h_j$ ($j=1,\dots ,N$), is 
\begin{equation}
\frac{\upd h_j}{\upd t}= K_j^{(1)}(\bH)+\eta_j(t)\;.
\end{equation}
The noise is Gaussian white, {\sl i.e.} with the mean value equal to zero and the covariance
\begin{equation}
\langle\eta_j(t)\eta_k(t')\rangle=K_{jk}^{(2)}\delta(t-t')\;.
\end{equation} 
We can separate the first transition moment $K_j^{(1)}$ into three summands, each of them taking into account a different process: deposition $F$ (or incoming flow), diffusion $D_j$, and/or volume conservation $C_j$, 
\begin{equation}
K_j^{(1)}=F +D_j+C_j\;,
\end{equation}
where 
\begin{eqnarray}
F\!&=&\!\!\frac{a}{\tau}\;,\\
D_j\!\!&=&\!\!\frac{a}{2\tau}\Bigl\lbrace\bigl[\Theta(h_{j+2}-h_{j+1})
-\Theta(h_j-h_{j+1})\bigr]-\bigl[\Theta(h_j-h_{j-1})-\Theta(h_{j-2}-h_{j-1})
\bigr]\Bigr\rbrace\;,\\
C_j\!\!&=&\!\!-\frac{a}{2\tau}\Bigl[\Theta(h_{j+2}-h_{j+1})\Theta(h_j-h_{j+1})-2\,\Theta(h_{j+1}-h_j)\Theta(h_{j-1}-h_j)
+\Theta(h_j-h_{j-1})\Theta(h_{j-2}-h_{j-1})\Bigr]\;.\nonumber\\
\end{eqnarray}
The diffusion term can be rewritten as
\begin{equation}
D_j=\frac{1}{2a}\Delta^{\!(1)}_2\cZa(\srho{-},\srho{+})\biggr\rfloor_{\srho{\pm}=(h_{j\pm 1}-h_j)/a}\,,\label{K1a}
\end{equation}
where $\Delta^{\!(1)}_2$ is the first variation between the NN columns [{\sl i.e.} $\Delta^{\!(1)}_2 F_j=F_{j+1}-F_{j-1}$] with the property $(2\,a)^{-1}\Delta^{\!(1)}_2\to\nabla$ when $a\to 0$. Here and below, we use the following notation for the partial derivative $\nabla\doteq \partial/\partial x$. The kernel $\cZa\!:\mathbb{R}^2\to\mathbb{R}$ is interpreted as a generalized function defined by  ($\srho{\pm}\!\in\mathbb{R}$)
\begin{equation}
\cZa(\srho{-},\srho{+})=\zeta\left[\Theta(\srho{+})-\Theta(\srho{-})\right]\;,\label{cZa}
\end{equation}
with $\zeta=a^2/\tau$. Notice that $\cZa$ is antisymmetric, {\sl i.e.} $\cZa(x,y)=-\cZa(y,x)$. The volume conserving term can be rewritten as
\begin{equation}
C_j=\frac{1}{a^2}\Delta^{\!(2)}_1\cZs(\vm,\vp)\biggr\rfloor_{\vpm=(h_{j\pm 1}-h_j)/a}\,,\label{K1s}
\end{equation}
where $\Delta^{\!(2)}_1$ is the second variation [{\sl i.e.} \mbox{$\Delta^{\!(2)}_1f_j=f_{j+1}-2\,f_j+f_{j-1}$}], with the property that $a^{-2}\Delta^{\!(2)}_1\to\nabla^2$ when $a\to 0$. Here, the kernel $\cZs\!:\mathbb{R}^2\to\mathbb{R}$ is the generalized function defined by
\begin{equation}
\cZs(\srho{-},\srho{+})=-\frac{a\,\zeta}{2}\;\Theta(\srho{-})\,\Theta(\srho{+})\;.\label{cZs}
\end{equation}
Notice that $\cZs$ is symmetric, {\sl i.e.} $\cZs(x,y)=\cZs(y,x)$.
The term $C_j$ can be obtained by changing $\srho{\pm}\to -\srho{\pm}$ in the first moment of the Krug model with opposite sign \cite{Krug-97}. The real-valued generalized functions $\cZa$ and $\cZs$ can be related to real analytic functions $Z_\mathrm{a}$ and $Z_\mathrm{s}$, respectively, by regularization techniques. In the limit $a\to 0$ the diffusion term [eq.~(\ref{K1a})] $D_j\to\nabla\cdot Z_\mathrm{a}$ and the volume conserving term  [eq.~(\ref{K1s})] $C_j\to\nabla^2 Z_\mathrm{s}\,$. In the next section we show how the symmetric kernel of the  RDSR model renormalizes to zero ({\sl i.e.} $Z_\mathrm{s}= 0$).

\section{Continuous Langevin equation {\sl via} regularization\label{sec:regulariz}}

We use a slightly different procedure than other authors to obtain the same coefficients and equations \cite{Vvedensky-03}. In contrast to their work, we regularize the antisymmetric and symmetric kernels [eqs.~(\ref{cZa}) and (\ref{cZs}), respectively] and not the entire first transition moment at once. Using the regularization procedure, the Heaviside function $\Theta(x)$ can be replaced by a smooth real-valued function $\theta_\varepsilon(x)$ depending on the continuous parameter $\varepsilon$. The regularizing function satisfies $\theta_\varepsilon(n)\to\Theta(n)$ when $\varepsilon\to 0^+$ for all $n\in\mathbb{Z}$. There are several proposals to represent $\Theta$ for a shifted analytic function, including the following 
\begin{equation*}
\theta_\varepsilon(x)=\frac{1}{2}\int_{-\infty}^x\;\Bigl[\erf\Bigl(\frac{s+1}{\varepsilon}\Bigr)-\erf\Bigl(\frac{s}{\varepsilon}\Bigr)\Bigr]\upd s\;,
\end{equation*}
with $\varepsilon >0$, introduced in refs.~\cite{Haselwandter-06,Haselwandter-07} . 
The kernels $\cZs$ and $\cZa$ [eqs.~(\ref{cZs}) and (\ref{cZa}), respectively] can be regularized using the $\varepsilon$-theta function $\theta_\varepsilon$, {\sl i.e.}
\begin{eqnarray}
&&\cZa^{(\varepsilon)}(\srho{-},\srho{+})=\zeta\,\bigl[
\theta_\varepsilon(\srho{+})-\theta_\varepsilon(\srho{-})\bigr]\nonumber\\
&&\cZs^{(\varepsilon)}(\srho{-},\srho{+})=-\frac{a\,\zeta}{2}\;\theta_\varepsilon(\srho{+})\,\theta_\varepsilon(\srho{-})\;,\nonumber
\end{eqnarray}
with $\zeta=a^2/\tau$. Expanding the $\varepsilon$-theta in Taylor series around $x=0$
\begin{equation}
\theta_\varepsilon(x)=\sum_{k=0}\,\frac{A_k^{(\varepsilon)}}{k!}\,x^k\;,\label{e-theta}
\end{equation}
we find the following expansions [superscripts $(\varepsilon)$ of 
$A_k^{(\varepsilon)}$ are omitted hereafter]
\begin{eqnarray}
&&\cZa^{(\varepsilon)}(\srho{-},\srho{+})=\zeta\,\bigl(\srho{+}
-\srho{-}\bigr)\Bigl[A_1+\frac{1}{2}\,A_2\bigl(\srho{+}+\srho{-}\bigr)
+\frac{1}{3!}\,A_3\bigl(\srho{+}^2+\srho{+}\srho{-}+\srho{-}^2\bigr)
+\Or(3)\Bigr]\nonumber\;,\\
&&\cZs^{(\varepsilon)}(\srho{-},\srho{+})=-\frac{a\,\zeta}{2}\,\Bigl[A_0^2+A_0 A_1 \bigl(\srho{+}+\srho{-}\bigr)
+\frac{A_0 A_2}{2} \bigl(\vm^2-2\gamma\,\vm\vp+\vp^2\bigr)+\Or(3)\Bigr]\label{cetas}
\end{eqnarray}
where 
\begin{equation}
\gamma=-\frac{A_1^2}{A_0 A_2}\,.\label{gamma}
\end{equation}
Evaluating eqs.~(\ref{cetas}) we obtain
\begin{eqnarray}
&&\cZa^{(\varepsilon)}(\vm,\vp)\Bigr\rfloor_{\vpm=(h_{j\pm 1}-h_j)/a}=2\,\zeta\,L_j^{(2)}\Bigl[
A_1+\frac{A_2}{2}\,a\,L_j^{(1)}+\frac{A_3}{3!}\,N_j^{(-1/2)}+\Or(3)\Bigr]\;,\nonumber\\
&&\cZs^{(\varepsilon)}(\vm,\vp)\Bigr\rfloor_{\vpm=(h_{j\pm 1}-h_j)/a}=-\frac{a\,\zeta}{2}\,\Bigl[A_0^2+A_0 A_1\,a\,L_j^{(1)}+A_0A_2(1+\gamma)\,N_j^{(\gamma)}+\Or(3)\Bigr]
\end{eqnarray}
where the linear and quadratic terms are
\begin{eqnarray}
L_j^{(2)}&=&\frac{h_{j+1}-h_{j-1}}{2\,a}\;,\nonumber\\
L_j^{(1)}&=&\frac{h_{j+1}-2 h_j+h_{j-1}}{a^2}\;,\label{discrete}\\
N_j^{(\beta)} \!&=&\! \frac{(h_{j+1}-h_j)^2-2\,\beta\, (h_{j+1}-h_j)(h_{j-1}-h_j)+(h_{j-1}-h_j)^2}{2 \,a^2(\beta+1)}\;,\nonumber
\end{eqnarray}
with $-1<\beta\le 1$ the discretization parameter of the discretized nonlinear terms \cite{Buceta-05}. The usual choice $\beta=1$, called standard or post-point discretization, depends only on the height of the NN columns and, thus, the error of approximating $(\nabla h)^2$ is minimized. In contrast, the choice $\beta=0$, called anti-standard or prepoint discretization, corresponds to the arithmetic mean of the squared slopes around the interface sites. In this work, $\gamma$ depends on the coefficients of regularization through eq.~(\ref{gamma}). If $0\le\gamma\le 1$, for $A_0>0$ then $A_2<0$, these coincides with the result obtained in ref.~\cite{Jung-99}. The unusual choice $\beta=-1/2$, which has no clear discrete geometric meaning, has a continuous limit as we show below. Expanding eqs.~(\ref{discrete}) around $x=ja$, the discretized terms and their limits when $a\to 0$ are
\begin{eqnarray}
L_j^{(2)}&=&\nabla h+\frac{1}{16}\nabla^3 h\;a^2+\Or(4)
\quad\longrightarrow\quad\nabla h\;,\nonumber\\
L_j^{(1)}&=&\nabla^2 h+\frac{1}{12}\nabla^4 h\;a^2+\Or(4)
\quad\longrightarrow\quad\nabla^2 h\;,\label{continuous}\\
N_j^{(\beta)} &=&(\nabla h)^2+\frac{1}{4}\biggl(\frac{1-\beta}{1+\beta}\biggr)(\nabla^2 h)^2\;a^2+\Or(4)
\quad\longrightarrow\quad(\nabla h)^2\;.\nonumber
\end{eqnarray}
Notice that the limit of $N_j^{(\gamma)}$ does not depend on the discretization parameter $\gamma$, as shown in ref.~\cite{Buceta-05}. 
Taking $\zeta$ finite, the symmetric and antisymmetric contributions of eq.~(\ref{cLE}), respectively, are 
\begin{eqnarray}
&&Z_\mathrm{a}=\lim_{a\to 0}\cZa^{(\varepsilon)}(\vm,\vp)\Bigr\rfloor_{\vpm=(h_{j\pm 1}-h_j)/a}=2\,\zeta\Bigl[A_1+\frac{A_3}{3!}\,\abs{\nabla h}^2+ \Or(4)\Bigr]\,\nabla h\;,\nonumber\\ &&Z_\mathrm{s}=\lim_{a\to 0}\cZs^{(\varepsilon)}(\vm,\vp)\Bigr\rfloor_{\vpm=(h_{j\pm 1}-h_j)/a}=0\;.
\end{eqnarray}
Thus the continuous stochastic differential equation for $h\!=\!h(x,t)$ is the eq.~(\ref{IDe}) with
\begin{eqnarray}
&&\nu_0=2\,\zeta\,A_1\;,\label{regul-coef1}\\
&&\nu_2=\zeta\,A_3\;.\label{regul-coef2}
\end{eqnarray}
These coefficients depend on the chosen regularization and the regularization parameter $\varepsilon$ through the coefficients of the $\theta_\varepsilon$ [see eq.~(\ref{e-theta})]. It is important to take into account that the expansion in eq.~(\ref{IDe}) still includes terms of all orders, but in its coarse­-grained limit we find the EW equation~(\ref{EWe}) with the diffusion coefficient $\nu_0$ given by eq.~(\ref{regul-coef1}).

\section{Framework based in generalized functions theory\label{sec:framework}}

\begin{figure}
\begin{center}
\includegraphics[scale=.5]{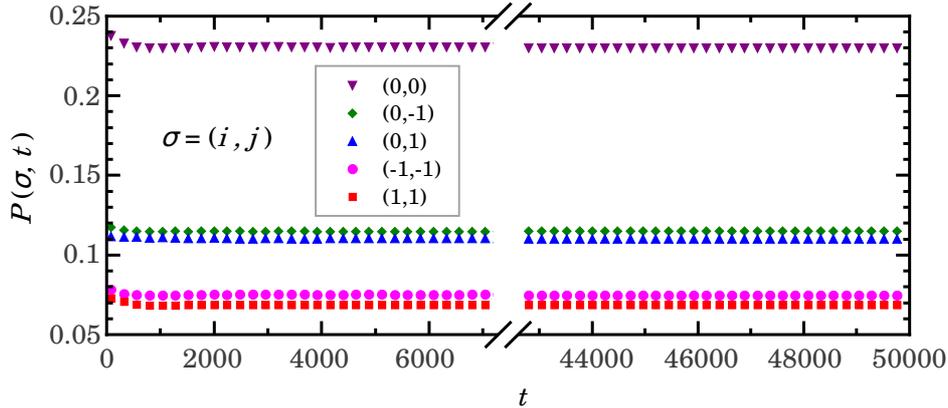}
\end{center}
\caption{(Color online) Plot of the probability density function $P(\bsigma,t)$ (PDF) as a function of time $t$ for various values ​​of $\bsigma$ indicated inside. The symbols show results of Monte Carlo simulations for the RDSR model with periodic boundary conditions and a lattice size $L=1024$ averaging over 25000 realizations. The broken axis is used in order to show the stationary behavior.\label{fig:1}}
\end{figure}

\begin{figure}
\begin{center}
\includegraphics[scale=.5]{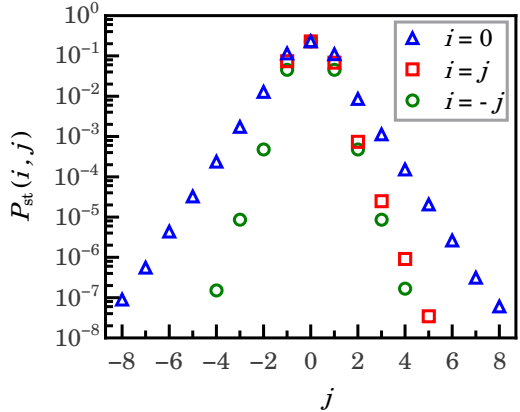}
\end{center}
\caption{(Color online) Semilog plot of the stationary probability density function $P_\text{st}(\bsigma)$ (SPDF) for different values ​​of $\bsigma=(i,j)$. Notice the symmetry of the SPDF (circles). Qualitatively, we observe in all cases that the data of SPDF show clear exponential decay.\label{fig:2}}
\end{figure}

According to the previous section, the kernels $\cZs$ and $\cZa$ are generalized functions of $\bvarrho=(\varrho_-,\varrho_+)\in\mathbb{R}^2$. In order to calculate statistical observables, at time $t$ it is possible to define a test function set $\varphi(\bvarrho,t)$ on which generalized functions $\cZ(\bvarrho)$ can be applied. The test functions are supported by the probability density function $P(\bsigma_j,t)$ to find any column $j$ with surface configuration $\bsigma_j\in\mathbb{Z}^2$ at time $t$ [{\sl i.e.} $\varphi(\bvarrho\!=\!\bsigma_j,t)=P(\bsigma_j,t)$], with $\bsigma_j=(\sigma_{j-},\sigma_{j+})$ and $\sigma_{j\pm}=(h_{j_\pm 1}-h_{j})/a$. These test functions  $\varphi$ must have different properties for restricted and unrestricted processes \cite{Buceta-12}. Unrestricted processes, {\sl e.g.} RDSR, require that $\varphi\in\mathcal{S}(\mathbb R^2)$, where $\mathcal{S}$ is the test space of $\mathrm{C}^\infty$-functions that decay and have derivatives of all orders that vanish faster than any power of $\varrho_\pm^{-1}$. {\sl Via} Monte Carlo simulations of the RDSR model, we show that $P(\bsigma_j,t)$ converges rather fast to stationary probability density function $\Pst(\bsigma_j)$ (SPDF). Figure~\ref{fig:1} shows that, for several configurations $\bsigma_j$, the $P(\bsigma_j,t)$ have a very long stationary regime which justifies the introduction of the time-independent test functions. We show also that the SPDF has at least exponential decay [See Figure \ref{fig:2}] which ensures $\varphi\in\mathcal{S}$. We define the test function $\varphi(\bvarrho)$ as a real-valued function supported by the discrete SPDF, {\sl i.e.} $\varphi(\bvarrho=\bsigma_j)=\Pst(\bsigma_j)$. We use here the notation on distributions that was introduced by Schwartz \cite{Schwartz-66}. 

The distribution $\cZ\in\mathcal{S}'$ (dual space of $\mathcal{S}$) applied to test function $\varphi\in \mathcal{S}$ is defined by
\begin{equation}
\langle\cZ\,,\varphi\rangle=\int_{\mathbb{R}^2}\cZ(\bvarrho)\,\varphi(\bvarrho)\;\mathrm{dv}_{\!\varrho}\;.\label{distr_f}
\end{equation}
Here $\langle Z,\varphi\rangle$ is the expectation value of $Z$ using the test function $\varphi$ as real-valued analytic representation of the SPDF. The test function is normed, {\sl i.e.} $\langle 1,\varphi\rangle=1\,$. The translation $T_{\balpha}$ of a distribution $\cZ$, denoted $T_{\balpha}\cZ$, extends the definition given by eq.~(\ref{distr_f}) to
\begin{equation}
\langle T_{\balpha}\cZ\,,\varphi\rangle=\langle\cZ\,, T_{-\balpha}\,\varphi\rangle\,,
\end{equation}
where the translation operator is defined by $T_{\bx}:\by\mapsto\by -\bx$ if $\by\,,\bx\in\mathbb{R}^2$ \cite{Colombeau-84}. As mentioned above, we assume that the test function $\varphi$ takes ​​fixed values ​​in the discrete lattice $\mathbb{Z}^2$ given by the SPDF $\Pst$, {\sl i.e.} $\langle T_{\bsigma}\delta\,,\varphi\rangle=\varphi(\bsigma)=\Pst(\bsigma)\,$ for all $\bsigma\in\mathbb{Z}^2\,$, where $\delta$ is the Dirac distribution. Applying a translation $T_{\bu}$ to a point $\bvarrho\in\mathbb{R}^2$, the transformation of test function is $\varphi\rightarrow T_{\bu}\varphi$\, if\,  $(\bvarrho-\bu)\in\mathbb{R}^2$. The change of the expectation value of $\cZ$ is $z(0)\rightarrow z(\bu)$ with
\begin{equation}
z(\bu)=\langle\cZ\,,T_{\bu}\varphi\rangle=\int_{\mathbb{R}^2}
\cZ(\bvarrho)\;\varphi(\bvarrho-\bu)\;\mathrm{dv}_{\!\varrho}\;.
\label{eq:w-u}
\end{equation}
For small translations, the Taylor expansion of $\varphi(\bvarrho-\bu)$ around $\bu =\boldsymbol 0\,$ is
\begin{equation}
\varphi(\bvarrho-\bu)=\varphi(\bvarrho)-u_\alpha\;\partial_\alpha\varphi\bigr\rfloor_{{\bu} =\mathbf{0}}+\tfrac{1}{2}\;u_\alpha u_\beta\;\partial^2_{\alpha\beta}\varphi\bigr\rfloor_{{\bu} =\mathbf{0}}+\Or(3)\;.\label{Tvarphi}
\end{equation}
Here repeated subscripts imply sum. The $\alpha$-th component of $\bu$ is $u_\alpha$. We used the notations $\partial_\alpha\,\dot=\,\partial/\partial\varrho_\alpha$ and $\partial^2_{\alpha\beta}\,\dot=\,\partial^2/(\partial\varrho_\alpha\partial\varrho_\beta)$, with $\alpha\,,\beta=1,2$. 
Since the test function $\varphi$ is known only at points of the lattice $\mathbb{Z}^2$, its derivatives cannot be calculated explicitly. In contrast, the distribution $\cZ$ is derivable in all points. Since the test function has either compact support or decreases rapidly, one can take advantage of the following identity 
\begin{equation}
\bigl\langle \cZ\,,\partial^{\rm{n}}_{\alpha\beta\cdots\omega}\varphi\bigr\rangle=(-1)^{\rm{n}}\bigl\langle \partial^{\rm{n}}_{\alpha\beta\cdots\omega} \cZ\,,\varphi\bigr\rangle\;.\label{repl}
\end{equation}
Using eq.~(\ref{repl}) the observable given by eq.~(\ref{eq:w-u}) for small translations is
\begin{equation}
z(\bu)=\Bigl\langle\cZ,\varphi\Bigr\rangle +\Bigl\langle\partial_\alpha\cZ,\varphi\Bigr\rangle\,u_\alpha
+\tfrac{1}{2}\Bigl\langle\partial^2_{\alpha\beta}\cZ,\varphi\Bigr\rangle\,u_\alpha u_\beta+\Or(3)\,.\label{distr-exp0}
\end{equation}
In the original work, Buceta and Hansmann \cite{Buceta-12} studied a restricted and an unrestricted discrete model belonging to the KPZ universality class: the restricted solid-on-solid model and the ballistic deposition model, respectively. They used the test space translations to determine the coarse-grained coefficients of the continuous stochastic differential equation known as KPZ equation. They derived their coefficients from the transformed average velocity of the interface $v(\bu)=\langle K^{(1)},T_{\bu}\varphi\rangle$, where the first transition moment $K^{(1)}$ is the drift of corresponding discrete Langevin equation. In a previous work \cite{Hansmann-13}, they studied two volume conserving surface (VCS) models without restrictions and with conserved noise, which differ from each other in the symmetry of their dynamic hopping rules. In contrast to the original work, these two models allow to calculate analytically another observable quantity, which is the nonconserved noise intensity. Thus, the continuous noise intensity is $D=\langle K^{(2)},\varphi\rangle$, where $K^{(2)}$ is the second transition moment corresponding to discrete process. The working methodology introduced to study the symmetric and asymmetric VCS models is used here to study the RDSR model. 

\section{Continuous Langevin equation coefficients {\sl via} test functions\label{sec:coef-test}}

Instead of finding the continuous Langevin equation {\sl via} regularization, we can obtain its non-zero coefficients $\nu_0$ and $\nu_2$ applying the distribution $\cZa$ [Eq.~(\ref{cZa})] on translated test functions $T_{\bu}\varphi$. 
In order to calculate the coefficients of the antisymmetric kernel $Z_\mathrm{a}$ [Eq.~(\ref{kernel-Za})] we perform a translation $\bu=(s,-s)$ on the surface configuration space with $s\ll 1$, taking into account that $\varphi$ is symmetric in its variables
\begin{equation}
z_\mathrm{a}(s) = \langle\cZa\,,T_{\bu}\varphi\rangle = \nu_0\,s+\frac{\nu_2}{3}\,s^3+\Or(s^5)\;,\label{za-kernel}
\end{equation} 
where
\begin{eqnarray}
\nu_0 &=& \bigl\langle(\partial_x-\partial_y)\cZa\,,\varphi\bigr\rangle\;,\label{nu_0}\\
\nu_2 &=& \frac{1}{2}\,\bigl\langle(\partial_x-\partial_y)^{\!(3)}\!\cZa\,,\varphi\bigr\rangle \label{nu_2}\;,
\end{eqnarray}
where $(\partial_x-\partial_y)^{\!(3)}=\partial_{xxx}^{\,3}-3\,\partial_{xxy}^{\,3}
+3\,\partial_{xyy}^{\,3}-\partial_{yyy}^{\,3}\;$.
Eq.~(\ref{za-kernel}) contains only odd powers in $s$ since $\cZa$ is antisymmetric and contributes only to their odd order derivatives which are symmetrical. Here we show only the first and third order-term, although the calculations can be extended to higher orders. From eq.~(\ref{cZa}) we obtain the symmetric distribution
$(\partial_x-\partial_y)\cZa=\zeta\bigl[\delta(x)+\delta(y)\bigr]$, hence the first-order coefficient (\ref{nu_0}) is
\begin{equation}
\nu_0 = 2\,\zeta\int_{-\infty}^{+\infty}\varphi(0,y)\,\upd y\;.\label{nu-0-i}
\end{equation}
In eq.~(\ref{nu_2}), taking into account the identity given by eq.~(\ref{repl})
$\nu_2=\tfrac{1}{2}\,\bigl\langle(\partial_x-\partial_y)\cZa\,,(\partial_x-\partial_y)^{\!(2)}\varphi\bigr\rangle$, the non zero terms contain $\bigl\langle\delta(x),\partial_{xx}^2\varphi\bigr\rangle=\bigl\langle\delta(y),\partial_{yy}^2\varphi\bigr\rangle$. Then, the third-order coefficient (\ref{nu_2}) is
\begin{equation}
\nu_2 =\zeta\,\int_{-\infty}^{+\infty}\partial_{xx}^2\varphi(x,y)
\Bigr\rfloor_{x=0}\;\upd y\;. \label{nu-2-i}
\end{equation}
The regularizing coefficients are related unequivocally with the test functions for the RDSR model in a simple manner [equalling eqs.~(\ref{regul-coef1}) and (\ref{nu-0-i}), and equalling eqs.~(\ref{regul-coef2}) and (\ref{nu-2-i})]. The coefficients are derived from the antisymmetric kernel, which  in this model is a linear combination of the $\Theta$-functions. For nonlinear combinations this relation between the regularizing coefficients and the test functions is unclear.  
\begin{figure}
\begin{center}
\includegraphics[scale=.55]{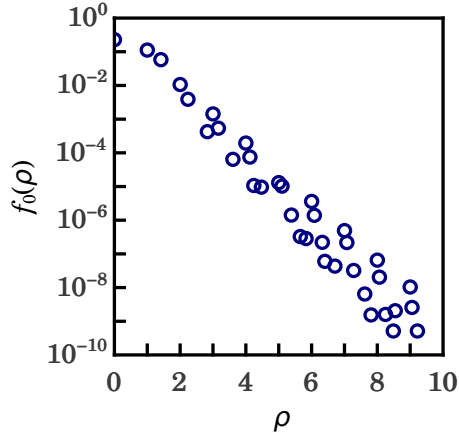}
\end{center}
\caption{(Color online) Semilog data plot of $f_0(\rho)$ obtained from the SPDF by means of eqs.~(\ref{sl1}), (\ref{sl4a}), (\ref{sl3a}) and (\ref{sl2}). We observe that the data of $f_0$ show exponential decay.\label{fig:3}}
\end{figure}

As the test function $\varphi(x,y)$ is symmetric under the exchange of its variables, it admits the following Fourier-cosine expansion
\begin{equation}
\varphi(x,y)=\sum_{n=0}^{+\infty}\,f_n(\rho)\,\cos[n(\theta-\pi/4)]\;,\label{test-fourier}
\end{equation}
which is an invariant function under the exchange $\theta\leftrightarrow\pi/2-\theta$. Considering the fact that the test function is known at those points where the SPDF is defined, the Fourier coefficients $f_n(\rho)$ can be easily determined on those points. Replacing the Fourier series given by eq.~(\ref{test-fourier}) in the normalization condition
\begin{equation}
\iint_{\mathbb{R}^2}\varphi(x,y)\,\upd x\upd y=1\;,
\end{equation}
we obtain the following equivalent condition for the zero-order Fourier coefficient
\begin{equation}
\int_0^{+\infty}\rho\;f_0(\rho)\,\upd\rho=1\;.
\end{equation}
The coefficients given by eqs.~(\ref{nu-0-i}) and (\ref{nu-2-i}) can be expressed in terms of the Fourier coefficients of eq.~(\ref{test-fourier}). See Appendix B for details. The lowest order contributions are
\begin{eqnarray}
\nu_0\simeq 4\,\zeta\int_0^{+\infty}f_0(\rho)\,\upd\rho\;,\label{nu-0-f}\\
\nu_2\simeq 32\,\zeta\int_0^{+\infty}\frac{f_4(\rho)}{\rho^2}\,\upd\rho\;,\label{nu-2-f}
\end{eqnarray}
It is possible to numerically compute some of these coefficients from simulation data of SPDF $P_\text{st}(i,j)=\varphi(i,j)$. In Appendix \ref{apA} we calculate the radial functions $f_0$ and $f_4$ given by the SPDF. Figure \ref{fig:3} shows the values of $f_0$ from Monte Carlo simulation data. 

\section{Summary and Conclusions\label{sec:concl}}

We have revisited a random deposition model with surface relaxation which belongs to the Edwards-Wilkinson (EW) universality class. We have established a general framework that can be extended to other discrete models such as molecular beam epitaxial growth models. We separated the different processes of the RDSR model into deposition, diffusion and volume conservation. This separation revealed the symmetries involved in the growth process and simplified further calculations. We explained that only the antisymmetric contributions determine the characteristic diffusion related to the EW universality class. In addition, we showed that only the symmetric contributions are associated to the volume conservation, which in the case of RDSR model renormalized to zero at the continuous limit. We led the RDSR model to the continuum in two different ways to obtain both, the continuous Langevin equation (or their coefficients) and the EW equation at the coarse-grained limit. For the first way, we used a regularization approach of generalized functions. We regularized the symmetric and antisymmetric kernels, which are nested in the first transition moment, in order to find the continuous equation. For the second way, we applied the generalized functions on test functions in order to calculate the coefficients of the continuous equation. The latter approach has the advantage that the coefficient could be estimated from SPDF data of a Monte Carlo simulation. The disadvantage of the approach is that the set of test functions, which is defined by the SPDF, makes the calculation of the coefficients imprecise. In contrast, the first approach has the two disadvantages that its results depend on the chosen regularizing function and the impossibility to calculate the coefficients from the simulation data. In general, the connection between the two approaches is not yet entirely clear. Up to now is only understandable in the RDSR model treated here, where the regularized function is a linear combination of Heavisides. In most other cases, the regularization of a generalized function (product of Heavisides) is not equal to the product of regularized Heavisides \cite{Oberguggenberger-92}. Finally, we discuss the applicability of our methodology to other discrete models. This formalism can be extended to other models with deposition and relaxation rules and nonconserved noise, such as Wolf-Villain MBE model \cite{Wolf-90} in 1+1 dimensions. This model reaches the steady-state after a very long time and studies using dynamic renormalisation group theory show that it asymptotically belongs to the EW universality class \cite{Haselwandter-07}. There is a very long transient, during which the system goes through different universality classes. This suggests that the coefficients of the continuous stochastic differential equation are time-dependent functions. These coefficients change their behaviour substantially when the universality class changes. Additionally, the PDF of height differences also shows a long transient. Therefore, following our approach, the test functions and coefficients are also time-dependent. Furthermore, this methodology is applicable to other discrete models with deposition and relaxation belonging to the EW universality class \cite{Pal-99}. Based on our work, we believe that the approaches discussed in this paper would facilitate similar research on other discrete stochastic models which involve deposition processes followed by instantaneous relaxation.

\section*{Acknowledgements}
D.H. gratefully acknowledges CONICET and MINCyT for their support.

\section*{Appendix A\label{apA}}

\begin{figure}
\begin{center}
\includegraphics[scale=.3]{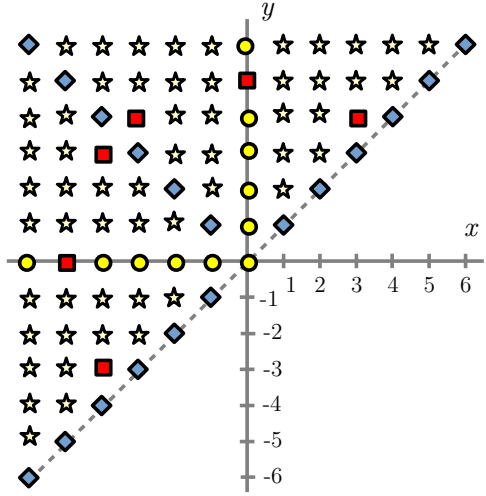}\hspace{1cm}
\end{center}
\caption{(Color online) Symbols show the integer coordinates $(i,j)$ where the test function $\varphi$ is known, {\sl i.e.} $\varphi(i,j)=P_\text{st}(i,j)$. We show a half-space plot, taking into account the symmetry property of the test function. Each symbol type corresponds to an equation system for the calculus of all possibles values of the Fourier coefficients of eq.~(\ref{test-fourier}); yellow circle: eq.~(\ref{yellow}), red square: eq.~(\ref{red}), white star: eq.~(\ref{white}), and blue diamond: eq.~(\ref{blue}).\label{fig:4}}
\end{figure}

In this appendix we calculate all possible values of the Fourier coefficients of equation~(\ref{test-fourier}) in terms of the known values of the test function. See Figure~\ref{fig:4} for details. Explicitly, if $j\in\mathbb{N}_0-\Upsilon$ where $\Upsilon=\{\upsilon\in\mathbb{N}/\upsilon=\sqrt{k^2+\ell^2},\;\forall\; k,\ell\in\mathbb{N}^\ast\;\text{and}\; k>\ell \}$
\begin{equation}
\left(
\begin{array}{c}
\varphi(0,j)\\\varphi(-j,0)
\end{array}
\right)=
\left(
\begin{array}{cc}
1&\frac{1}{\sqrt{2}}\\
1&-\frac{1}{\sqrt{2}}
\end{array}
\right)\left(
\begin{array}{c}
f_0(j)\\f_1(j)
\end{array}
\right)\;,\label{yellow}
\end{equation}
and $f_n(j)=0$ for $n\ge 2$\,. By solving the system
\begin{equation}
\left(
\begin{array}{c}
f_0(j)\\f_1(j)
\end{array}
\right)=\frac{1}{2}\left(
\begin{array}{cc}
1&1\\
\sqrt{2}&-\sqrt{2}
\end{array}
\right)\left(
\begin{array}{c}
\varphi(0,j)\\\varphi(-j,0)
\end{array}
\right)\;.\label{sl1}
\end{equation}
Otherwise, with $j\in\Upsilon$, the second case is  
\begin{equation}
\left(
\begin{array}{c}
\varphi(\ell,k)\\\varphi(0,j)\\\varphi(-\ell,k)\\\varphi(-k,\ell)\\\varphi(-j,0)\\\varphi(-k,-\ell)
\end{array}
\right)=\left(
\begin{array}{cccccc}
1 &\cos\alpha &\cos 2\alpha &\cos 3\alpha &\cos 4\alpha &\cos 5\alpha\\
1 &\frac{1}{\sqrt{2}} &0 &-\frac{1}{\sqrt{2}} &-1 &-\frac{1}{\sqrt{2}}\\
1 &\sin\alpha &-\cos 2\alpha &-\sin 3\alpha &\cos 4\alpha &\sin 5\alpha\\
1 &-\sin\alpha &-\cos 2\alpha &\sin 3\alpha &\cos 4\alpha &-\sin 5\alpha\\
1 &-\frac{1}{\sqrt{2}} &0 &\frac{1}{\sqrt{2}} &-1 &\frac{1}{\sqrt{2}}\\
1 &-\cos\alpha &\cos 2\alpha &-\cos 3\alpha &\cos 4\alpha &-\cos 5\alpha
\end{array}
\right)\left(
\begin{array}{c}
f_0(j)\\f_1(j)\\f_2(j)\\f_3(j)\\f_4(j)\\f_5(j)
\end{array}
\right)\;,\label{red}
\end{equation}
where $j=\sqrt{k^2+\ell^2}$ ($k,\ell\in\mathbb{N}^\ast,\hspace{1ex}k>\ell$) and $\alpha=\arctan(k/\ell)-\pi/4$ (with $0<\alpha<\pi/4$). Solving the equation system
\begin{equation}
\left(
\begin{array}{c}
f_0(j)\\f_2(j)\\f_4(j)
\end{array}
\right)=\frac{1}{4\cos^2 2\alpha}
\left(
\begin{array}{ccc}
\frac{1}{2} &\cos 4\alpha &\frac{1}{2} \\
\cos 2\alpha &0 &-\cos 2\alpha\\
\frac{1}{2} &-1 &\frac{1}{2}
\end{array}
\right)\left(
\begin{array}{c}
\varphi(\ell,k)+\varphi(-k,-\ell)\\ \varphi(0,j)+
\varphi(-j,0)\\ \varphi(-\ell,k)+\varphi(-k,\ell)
\end{array}
\right)\;,\label{sl4a}
\end{equation}
\begin{equation}
\left(
\begin{array}{c}
f_1(j)\\f_3(j)\\f_5(j)
\end{array}
\right)=\frac{1}{2\,\sin 4\alpha\,(1+\cos 4\alpha)}
\!\!\left(\!\!
\begin{array}{ccc}
\sin 4\alpha\,\cos\alpha&\frac{1}{\sqrt{2}}\sin 8\alpha &\sin 4\alpha\,\sin\alpha\\
\sin 3\alpha\,\cos 2\alpha &-\frac{1}{\sqrt{2}}\sin 4\alpha  &-\cos 3\alpha\,\cos 2\alpha\\
\sin\alpha\,\cos 2\alpha  &-\frac{1}{\sqrt{2}}\sin 4\alpha &\cos\alpha \,\cos 2\alpha
\end{array}
\!\!\right)\!\!\left(\!
\begin{array}{c}
\varphi(\ell,k)-\varphi(-k,-\ell)\\ \varphi(0,j)-
\varphi(-j,0)\\ \varphi(-\ell,k)-\varphi(-k,\ell)
\end{array}
\!\right)\!\!.\label{sl4b}
\end{equation}
Also, if $\sigma\in\Sigma$ where $\Sigma=\{\sigma\in\mathbb{R}-\mathbb{N}/\sigma=
\sqrt{k^2+\ell^2},\;\forall\; k,\ell\in\mathbb{N}^\ast\;\text{and}\; k>\ell \}$
\begin{equation}
\left(
\begin{array}{c}
\varphi(\ell,k)\\\varphi(-\ell,k)\\\varphi(-k,\ell)\\\varphi(-k,-\ell)
\end{array}
\right)=\left(
\begin{array}{cccc}
1 &\cos\alpha &\cos 2\alpha &\cos 3\alpha\\
1 &\sin\alpha &-\cos 2\alpha&-\sin 3\alpha\\
1 &-\sin\alpha &-\cos 2\alpha &\sin 3\alpha\\
1 &-\cos\alpha &\cos 2\alpha &-\cos 3\alpha
\end{array}
\right)\left(
\begin{array}{c}
f_0(\sigma)\\f_1(\sigma)\\f_2(\sigma)\\f_3(\sigma)
\end{array}
\right)\;,\label{white}
\end{equation}
where  $\ell\neq k$ and $\alpha=\arctan(k/\ell)-\pi/4$ (with $0<\alpha<\pi/4$). By solving the system
\begin{equation}
\left(
\begin{array}{c}
f_0(\sigma)\\f_2(\sigma)
\end{array}
\right)=\frac{1}{4\cos 2\alpha}
\left(
\begin{array}{ccc}
\cos 2\alpha &\cos 2\alpha\\
1 &-1
\end{array}
\right)\left(
\begin{array}{c}
\varphi(\ell,k)\,+\,\varphi(-k,-\ell)\\ \varphi(-\ell,k)\,+\,\varphi(-k,\ell)
\end{array}
\right)\;,\label{sl3a}
\end{equation}
\begin{equation}
\left(
\begin{array}{c}
f_1(\sigma)\\f_3(\sigma)
\end{array}
\right)=\frac{1}{2\,\sin 4\alpha}
\left(
\begin{array}{ccc}
\sin 3\alpha &\cos 3\alpha\\
\sin\alpha &-\cos\alpha
\end{array}
\right)\left(
\begin{array}{c}
\varphi(\ell,k)\,-\,\varphi(-k,-\ell)\\ \varphi(-\ell,k)\,-\,\varphi(-k,\ell)
\end{array}
\right)\;.\label{sl3b}
\end{equation}
Finally, explicitly for $j\in\mathbb{N}_0$
\begin{equation}
\left(
\begin{array}{c}
\varphi(j,j)\\\varphi(-j,j)\\\varphi(-j,-j)
\end{array}
\right)=\left(
\begin{array}{ccc}
1 &1 &1\\
1 &0 &-1\\
1 &-1 &1
\end{array}
\right)\left(
\begin{array}{c}
f_0(\sqrt{2}j)\\f_1(\sqrt{2}j)\\f_2(\sqrt{2}j)
\end{array}
\right)\;,\label{blue}
\end{equation}
and $f_n(j)=0$ for $n\ge 3$\,. By solving the system
\begin{equation}
\left(
\begin{array}{c}
f_0(\sqrt{2}j)\\f_1(\sqrt{2}j)\\f_2(\sqrt{2}j)
\end{array}
\right)=\frac{1}{4}\left(
\begin{array}{ccc}
1 &2 &1\\
2 &0 &-2\\
1 &-2 &1
\end{array}
\right)\left(
\begin{array}{c}
\varphi(j,j)\\\varphi(-j,j)\\\varphi(-j,-j)
\end{array}
\right)\;.\label{sl2}
\end{equation}
 
\section*{Appendix B\label{apB}}

In order to obtain the coefficient $\nu_0$ in terms of Fourier coefficients [eq.~(\ref{test-fourier}], it is easy to show that the integral of eq.~(\ref{nu-0-i}) is
\begin{equation}
\int_{-\infty}^{+\infty}\varphi(0,y)\,\upd y=\sum_{n=0}^{+\infty}\bigl[1+(-1)^n\bigl]\cos\Bigl(\frac{n\pi}{4}\Bigr)\!\int_0^{+\infty}\!f_n(\rho)\,\upd\rho=2\,\sum_{k=0}^{+\infty}(-1)^k\!\int_0^{+\infty}f_{4k}(\rho)\,\upd\rho\;.\label{int-nu-0}
\end{equation}
Similarly, we obtain the coefficient $\nu_2$ in terms of Fourier coefficients [eq.~(\ref{test-fourier}]. Take into account
\begin{eqnarray}
&&\frac{\partial\hspace{1ex}}{\partial x}=\cos\theta\,\frac{\partial\hspace{1ex}}{\partial\rho}-\frac{1}{\rho}\,\sin\theta\,\frac{\partial\hspace{1ex}}{\partial \theta}\nonumber\\
&&\frac{\partial\hspace{1ex}}{\partial y}=\sin\theta\,\frac{\partial\hspace{1ex}}{\partial\rho}+\frac{1}{\rho}\,\cos\theta\,\frac{\partial\hspace{1ex}}{\partial \theta}\;, 
\end{eqnarray}
is direct to show
\begin{eqnarray*}
&&\left.\frac{\partial^2\varphi}{\partial x^2}\right\rfloor_{x=0,y>0}=-\frac{1}{y^2}\;\sum_{n=1}^{+\infty}n^2\cos\Bigl(\frac{n\pi}{4}\Bigr)f_n(y)\\
&&\left.\frac{\partial^2\varphi}{\partial x^2}\right\rfloor_{x=0,y<0}=-\frac{1}{y^2}\;\sum_{n=1}^{+\infty}n^2\cos\Bigl(\frac{3 n\pi}{4}\Bigr)f_n(-y)\;.
\end{eqnarray*}
Then the integral of eq.~(\ref{nu-2-i}) is
\begin{equation}
\int_{-\infty}^{+\infty}\partial^2_{xx}\varphi\Bigr\rfloor_{x=0}\upd y=-
\sum_{n=1}^{+\infty}n^2\bigl[1+(-1)^n\bigr]\,\cos\Bigl(\frac{n\pi}{4}\Bigr)\int_0^{+\infty}\frac{f_n(\rho)}{\rho^2}\,\upd\rho=-32\,\sum_{k=1}^{+\infty}(-1)^k\,k^2\!\int_0^{+\infty}\frac{f_{4k}(\rho)}{\rho^2}\,\upd\rho\;.\label{int-nu-2}
\end{equation}
The series of eqs.~(\ref{int-nu-0}) and (\ref{int-nu-2})  can be approximated to the lowest order taking into account that the integrals of the series are strongly convergent. 

%\section*{References}
\bibliography{competitive-Buceta-Hansmann}

\end{document}